\documentclass[12pt]{iopart}

\usepackage{iopams} 
\usepackage{graphicx}
\usepackage{cite}

\newif\ifextended

\extendedtrue

\begin{document}

\title[Comment on `Radiation from multi-GeV electrons and
positrons ...']{Comment on `Radiation from multi-GeV electrons and
positrons in periodically bent silicon crystal'}

\author{Andriy Kostyuk}

\address{Ostmarkstra{\ss}e 7, 93176 Beratzhausen, Germany}
\ead{andriy.p.kostyuk@gmail.com}

\begin{abstract} 
Simulations of electron and positron channelling 
in a crystalline undulator with a small amplitude and a short 
period (A~Kostyuk, Phys. Rev. Lett. 110 (2013) 115503) were 
repeated by V~G~Bezchastnov, A~V~Korol and A~V~Solovʼyov 
(J. Phys. B: At. Mol. Opt. Phys. 47 (2014) 195401)
for the same parameter set but using another computer code,
MBN Explorer, which implements
a different model of projectile scattering by crystal atoms.
The authors of the latter paper claim
that their approach, in contrast to the one of the former paper,
allows them to observe short-period undulator oscillations
in plots of simulated trajectories.
In fact,
the undulator oscillations become visible
on trajectory segments which have a small amplitude of channelling oscillations.
This is equally true for both approaches.
The claim of Bezchastnov \textit{et al.} that their model is 
\textit{``more accurate''} is unfounded. Moreover, there are indications of 
severe mistakes in their calculations:
the phases of undulator 
oscillations in the trajectories obtained with MBN Explorer are inconsistent 
with each other and with  well established properties of forced oscillations.
\end{abstract}

\maketitle

The authors of \cite{Bezchastnov2014} repeated numerical studies   
of electron and positron channelling in a small-amplitude short-period 
crystalline undulator (SASP CU) which initially had been done 
in \cite{Kostyuk2012}.
\ifextended
The simulations were performed at the same conditions as in the  
original paper: the projectile energy $E=855$ MeV channelled in 
a 12 $\mu$m thick Si crystal along the (110) plane 
sinusoidally bent with the period 
$\lambda_\mathrm{u}=400$ nm and the amplitude $a_\mathrm{u}=0.4$ \AA.
\fi
Bezchastnov \textit{et al.} were sceptical about the fact
that the short-period crystal bending can generate
a pronounced peak in the calculated radiation spectra
of channelled particles, although
its effect is not seen in the sample trajectories
plotted in \cite{Kostyuk2012}.
The authors of \cite{Bezchastnov2014} characterized
this result as \textit{``a puzzling if not a misleading one''}.
The necessity of the new study they justified as follows.
\begin{quote}
\textit{``To address the essence of undulator radiation it is
imperative to study whether the projectiles acquire the short-
period bound oscillations when passing through the CU.
Giving the positive answer leaves no puzzle in understanding
the spectral lines in CU radiation''.}
\end{quote}
It is stated in \cite{Bezchastnov2014} that
\ifextended
\textit{``to elucidate the properties of motion and
radiation in the CU''}
\fi
the authors
\textit{``employ the theoretical methods more
advanced and accurate that those in the studies''} \cite{Kostyuk2012}.
However, no description of any new theoretical methods is
present in \cite{Bezchastnov2014}. The authors reveal only that they use
another model for projectile scattering by crystal atoms.

The original calculations \cite{Kostyuk2012} were done with the computer code ChaS 
(Channelling Simulator). This code implements the \textit{snapshot model} of the crystal atoms 
based on Moli{\`er}e's potential \cite{Moliere1947}.
(The definition and a detailed description of the snapshot model can be found in 
\cite{Kostyuk2010,Kostyuk2011a,Kostyuk2011c}).
In contrast, the MBN (Meso-Bio-Nano) Explorer package \cite{Sushko2013a}
used in \cite{Bezchastnov2014}
calculated the projectile scattering by Moli{\`er}e's potential directly.
The authors of \cite{Bezchastnov2014} did obtain trajectories 
demonstrating visible oscillations induced by the crystal
bending. However, they did not provide any evidence that 
using the particular scattering model
\ifextended
or any other \textit{``advanced and accurate''} theoretical methods
\fi
was essential for obtaining this result.

It will be shown in the following that 
\begin{itemize}
 \item no puzzle regarding the (in)visibility of the undulator oscillations in the
 simulated trajectories has ever existed;
 \item undulator oscillations are well seen in plots of trajectories having 
 a small amplitude of channelling oscillations;
 \item such trajectories can be obtained with the code ChaS using exactly the same 
 simulation algorithm that was used in \cite{Kostyuk2012};
\ifextended
 \item there has been no reason to expect that the model used in \cite{Bezchastnov2014}
 would yield a larger undulator oscillation amplitude than the one of \cite{Kostyuk2012};
\fi
 \item the claim that the model of \cite{Bezchastnov2014} is more accurate than the one 
 of \cite{Kostyuk2012} is unfounded;
 \item the results of calculations published in \cite{Bezchastnov2014} are self-contradictory.
\end{itemize}

First of all, it has to be made clear that observing the effect of the crystal bending
in plots of simulated trajectories was not a goal of the study presented in \cite{Kostyuk2012}.
This question is not of substantial scientific interest because
the projectile trajectory is not observed in the experiment and cannot be compared to the simulation 
results. In contrast, the spectrum of the 
emitted radiation can be measured and compared to the one calculated using the simulated trajectories. 
Therefore, the determination of the shape of the radiated spectra from projectile electrons and 
positrons channelled in a SASP CU was the central objective of the study. On the other hand, 
it follows from the properties of Fourier's transform that
the spectrum contains all essential
information about the shape of the trajectories. In particular, the presence of a
pronounced peak at the corresponding frequency in the spectrum leaves no doubt that
the projectiles do acquire short-period oscillations when passing through a SASP CU.
Hence, \textit{``the positive answer''} the author of \cite{Bezchastnov2014} were looking
for had already been given in \cite{Kostyuk2012}.

\ifextended
The purpose of the sample trajectories that were plotted in \cite{Kostyuk2012} was 
not to demonstrate the undulator oscillations in their shape. 
They were plotted  
to illustrate two new features of SASP CU distinguishing it from the 
previously known large-amplitude long-period crystalline 
undulator (LALP CU). First, channelling in SASP CU is still possible 
despite the broken centrifugal condition (inequality (4) of \cite{Kostyuk2012}).
Second, the projectile does not follow the shape of the channel.
\fi

\ifextended
What the authors of \cite{Bezchastnov2014} call \textit{``conclussion''} was
a remark referring to the particular sample trajectories. Its formulation,
\textit{``it is practically impossible to see the
modification of the trajectories due to the crystal 
bending in Fig. 1''} (here figure 1 of \cite{Kostyuk2012} is meant),
was leaving no space for interpreting it as a general judgement.
\fi

The reason why the undulator oscillations are not seen in the plotted
trajectories has been clear as well. The amplitude $a$ of the undulator oscillations
is very small. It is much smaller than the amplitude $a_\mathrm{u}$ of the crystal bending,
while the amplitude $a_\mathrm{c}$ of channelling oscillations of the trajectories 
plotted in \cite{Kostyuk2012} is approximately equal or even slightly larger than the bending 
amplitude, therefore 
\begin{equation}
  a \ll  a_\mathrm{c}. \label{large_chan_osc}
\end{equation}
In other words, the shape of the trajectory is dominated by the channelling oscillations.
This is why  it is difficult to see the crystal bending effect in the plots.

The seemingly paradoxical fact that the undulator radiation peak in the radiation spectrum
(figures 4 and 5 of \cite{Kostyuk2012}) is higher than the channelling
radiation maximum despite of the smallness of the undulator oscillation amplitude
relative to the typical channelling oscillation amplitude 
was explained in \cite{Kostyuk2012,Kostyuk2013d}.
This happens because the intensity of radiation is proportional to the fourth power of the oscillation
frequency and the frequency of the undulator oscillations in the case of SASP CU is substantially 
lager than the one of the channelling oscillations. 

From the above arguments it must be clear that 
no puzzles regarding the  undulator oscillations in the simulated trajectories and the 
corresponding radiation peak have ever existed.

\begin{figure}[htb]
\begin{center}
\includegraphics[width=0.5\linewidth]{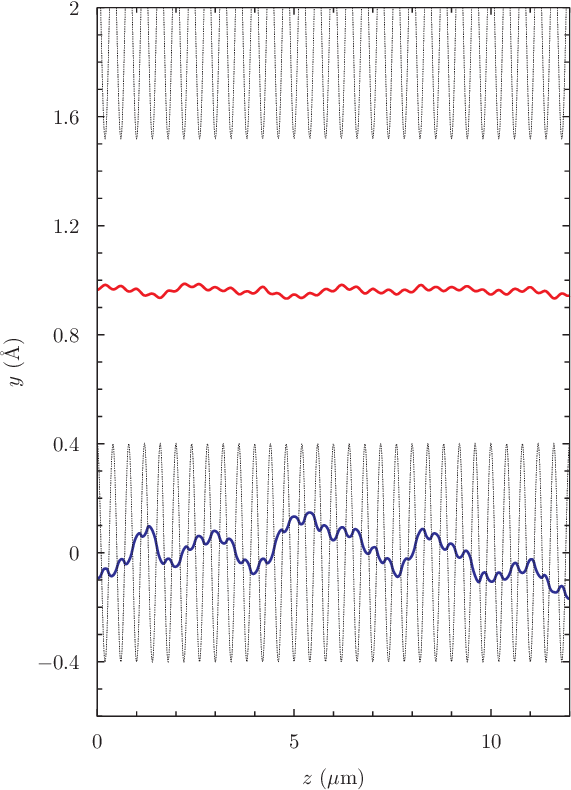}
\end{center}
\caption{Trajectories with a small amplitude of channelling oscillations of a positron (above)
and an electron (below) simulated with the code ChaS (Channelling Simulator). 
The simulations were done for projectiles with the energy $E=855$ MeV 
channelled along (110) plane of a silicon crystal
bent with the period 
$\lambda_\mathrm{u}=400$ nm and amplitude $a_\mathrm{u}=0.4$ \AA.
The trajectories were selected 
from two  ensembles of, respectively, positron and electron trajectories,  containing $\sim 10^7$ trajectories each.
The selection criterion was the smallest channelling oscillation amplitude within the 12 $\mu$m long trajectory.
The dashed wavy lines show the shape of the bent crystal planes.
Undulator oscillations are well seen and have an opposite phase with respect to the crystal bending.
\label{small_amplitude}}
\end{figure}

The strong inequality (\ref{large_chan_osc}) is valid for typical trajectories like those plotted
in  \cite{Kostyuk2012,Kostyuk2013d}. Still, one can select trajectories with
nearly vanishing channelling oscillations. Two examples of such trajectories, one for a positron
and one for an electron, are shown in figure \ref{small_amplitude}. As expected, the undulator oscillations
are well seen in the plot. 
It has to be stressed that the trajectories shown in figure \ref{small_amplitude}
were obtained with the same algorithm that was used in \cite{Kostyuk2012,Kostyuk2013d}.
If any \textit{``theoretical methods more advanced and accurate
 that those in the studies''} \cite{Kostyuk2012} were employed in
 \cite{Bezchastnov2014}, they were redundant.
Application of a different model of the projectile scattering was not essential either.
One sees in figure 1 of \cite{Bezchastnov2014} that the undulator oscillations are well seen on the trajectory 
fragments having nearly vanishing amplitude of channelling oscillation, exactly as in figure \ref{small_amplitude} of the present paper. 
In contrast, the effect of the crystal bending is barely seen  if the channelling oscillation amplitude becomes  
comparable with or larger than  the amplitude of the crystal bending, exactly as it was in   \cite{Kostyuk2012,Kostyuk2013d}.

\ifextended
Did the authors of \cite{Bezchastnov2014} have reasons to expect that employing the direct potential scattering model instead of the
snapshot model would increase the visibility of the undulator oscillations in the simulated trajectories? In fact,
they did not. In the article \cite{Sushko2013a}, that preceded \cite{Bezchastnov2014} and was cited there, its authors claimed 
\begin{quote}
\textit{``a ``snapshot'' approximation overestimates the mean scattering angle
in a single projectile---atom collision''.}
\end{quote}
If this statement were true,  the interaction of the projectile with the average
potential would be weaker than with a snapshot atom  and so would be the effect of the crystal bending in the simulated trajectories. 
Therefore, one would have to expect  that 
using the potential scattering model instead of  the snapshot model would make an observation of undulator oscillations 
in plots of simulated trajectories even less likely.  

 In reality, however, the above statement of \cite{Sushko2013a} is wrong.
  
For the above reason,  the average scattering
angle in the snapshot model \textit{equals} to the scattering angle in the potential  
$\varphi(\vec{r})$ for the same impact parameter.

The same is true for any other 
quantity that depends linearly on the potential or its derivatives. The amplitude 
of the undulator oscillations is linearly proportional to amplitude of the 
variation of the potential gradient along the trajectory. 
Therefore, this amplitude is the same in the snapshot model as in the 
direct potential scattering model.
Hence, there was no reason to expect that the direct potential scattering model
is suited better than the snapshot model 
for the observation of the undulator oscillations in a plot
of simulated trajectories.
\fi

The authors of \cite{Bezchastnov2014} claim that MBN Explorer
\textit{``is designed for high-accuracy computations''}.
Further they allege that their approach
\textit{``treats the forces experienced by the projectiles in 
a much more accurate way''} than the snapshot model.
They do not reveal, however, how they came to this
conclusion.
They refer to \cite{Sushko2013a} for details, but this paper does not 
contain any accuracy assessment of 
the forces acting on the projectile in 
the two models. 
However, the authors of \cite{Sushko2013a} evaluated
the average potential of the snapshot model
and came to the conclusion that it 
``reproduces the Moli{\`e}re potential quite accurately''
(see page S4 of the ``Supporting material'' of \cite{Sushko2013a}).
Moreover, the model implemented in the  MBN Explorer does not 
take into account incoherent scattering of the projectile by crystal 
electrons \cite{Kostyuk2018} which raises doubts about the attainability of 
the high-accuracy in the modeling of channeling. 

Besides of that, a thorough inspection of figure 1 in \cite{Bezchastnov2014} 
reveals inconsistencies in the phase shift of the undulator oscillations 
with respect to the crystal bending.
The frequency of the periodic force
acting onto the projectile due to the crystal bending exceeds substantially the frequency
of the channelling oscillations, which is the resonance frequency
(also known as \textit{eigenfrequency}) of the system.
In this case the phase of the forced oscillations has to be opposite  to the phase
of the periodic force (see e.g. \S 26 of \cite{Landau1976CM}). The phase of the
periodic force acting on a positron or
on an electron moving in the middle part of the bent crystal channel coincides with
the phase of the bending. Therefore, the phase of the undulator oscillations
has to be opposite to  the one of the crystal bending. One sees in
in figure \ref{small_amplitude} of the present paper that this is indeed the
case for the trajectories simulated with the code ChaS.
In contrast, the relative phase of the undulator oscillations of the
trajectories obtained using MBN Explorer (figure 1 of \cite{Bezchastnov2014})
changes along the crystal.
\ifextended
On some segments it coincides with the phase of
the crystal bending, in other ones it is opposite or has an intermediate
shift.
\fi
The period of the crystal bending shown with the dashed lines in
figure 1 of \cite{Bezchastnov2014} is somewhat larger than
the value of $\lambda_\mathrm{u}=400$~nm given in the figure
caption. One could attribute the phase mismatch of the undulator
oscillations to this discrepancy if the phase slip pattern were the same for 
all three trajectories. In fact, the two electron trajectories in 
figure 1 of \cite{Bezchastnov2014} have different if not opposite phases of undulator
oscillations, i.e. the results shown in figure 1 of \cite{Bezchastnov2014} are self-contradictory.
\ifextended
This is a clear indication that there must be one more mistake either
in the source code of MBN Explorer or in the plot preparation script.
\fi
Until the reason for this inconsistency is clarified all results obtained with MBN Explorer
\cite{Sushko2013a,Sushko2013,Polozkov2014,Polozkov2015,Sushko2015,Sushko2015b,Sushko2015c,Korol2016,Korol2017}
should be taken with a significant degree of caution. 

\ifextended
In conclusion, I would like to stress the importance of scientific integrity.
In particular, contribution of other colleagues to the field of research has to be 
credited appropriately.
\fi
In the first paragraph on page 2 of \cite{Bezchastnov2014} the authors claim
that the concept of the crystalline undulator was formulated in their papers
\cite{Korol1998,Korol1999} published in 1998--1999. In fact, this concept had been known since
1980 \cite{Baryshevsky1980,Kaplin1980a,Kaplin1980}.

\end{document}